# Integrated Information Management for TESLA

J. Bürger, L. Hagge, J. Kreutzkamp, K. Lappe, A. Robben
*Deutsches Elektronen-Synchrotron, DESY, Notkestraße 85, 22607 Hamburg, Germany*

Next-generation projects in High Energy Physics will reach again a new dimension of complexity. Information management has to ensure an efficient and economic information flow within the collaborations, offering world-wide up-to-date information access to the collaborators as one condition for successful projects. DESY introduces several information systems in preparation for the planned linear collider TESLA: a Requirements Management System (RMS) is in production for the TESLA planning group, a Product Data Management System (PDMS) is in production since the beginning of 2002 and is supporting the cavity preparation and the general engineering of accelerator components. A pilot Asset Management System (AMS) is in production for supporting the management and maintenance of the technical infrastructure, and a Facility Management System (FMS) with a Geographic Information System (GIS) is currently being introduced to support civil engineering. Efforts have been started to integrate the systems with the goal that users can retrieve information through a single point of access. The paper gives an introduction to information management and the activities at DESY.

## 1. INTRODUCTION

Globalization and information technology have changed the boundary conditions for successful enterprises. Today's critical success factors include high speed of innovation and time-to-market, high quality standards, economic business design and international presence. To cope with these conditions, enterprises increasingly rely on CAx-technologies (CAD, CAM, CAQ, ...), and they re-engineer their business with the goal of improving information exchange and enabling simultaneous engineering on a global scale. Organizations address the entire life cycle of their products to allow for feedback from installation and even operation back to product design for improved quality. Building an accelerator in a High Energy Physics laboratory faces the same boundary conditions as mentioned above, hence it should be investigated if HEP can benefit from the same measures.

Next-generation projects in HEP will once more set a new scale of complexity in collaboration. The planned Linear Collider TESLA [1][2] will be realized by a global collaboration with decentralized envisioning, engineering and manufacturing. For these purposes, the collaboration will need location-independent access to up-to-date information covering engineering, technical infrastructure, facilities, project management, controlling and more.

At Deutsches Elektronen-Synchrotron (DESY) an information management group has been created to improve the transparency and help to ensure the performance of such large-scale future projects.

As a first step, new information systems which support the life cycle of an accelerator are being introduced at DESY: a Requirements Management System (RMS) is in production for the TESLA planning group, and a Product Data Management System (PDMS) is in production for supporting the cavity preparation and the general engineering of accelerator components. A pilot Asset Management System (AMS) is in production for supporting the management and maintenance of the technical infrastructure, and a Facility Management System (FMS) and a Geographic Information System (GIS) are currently being customized to support the civil engineering. Work is going on to integrate the systems so that users can access combined information.

The goal is to finally provide an integrated information management which supports the entire life cycle of collider projects. The vision is that

- experts use specialized tools which are optimized to their specific tasks,
- the expert tools are tightly integrated with a supporting information system which stores, relates and manages the emerging information, like e.g. 3D-CAD with EDMS or architectural CAD with FMS, and
- a general-purpose Web interface is available as a single point of access for navigating and retrieving information from any of the information systems.

This paper gives an introduction to the information management activities for TESLA at DESY. The paper first introduces the key concepts of information management. It then describes how information management can support the different stages of an accelerator life cycle and presents the different information systems being introduced and used at DESY, before it finally summarizes the project status and the experience gained.

## 2. INFORMATION MANAGEMENT

### 2.1. Key Concepts.

Information Management is an increasingly popular term in computer science and among IT professionals, which is used with different meanings. Information management is commonly used for the planning, budgeting, control and exploitation of the information resources in an organization [3]. From a management point of view, information management is responsible for supplying all the resources in an organization efficiently and economically with the information they need to achieve their business goals [4].

Matching (?) the information people request with the information they actually need for their work, and with the information they offer to others is rather complex (cf. Figure 1). If for example an additional component has to be inserted into a beam-line, the project team *needs* to know the available space, i.e. the dimension of the gap between the neighbouring components into which it should be inserted. Out of habit, the team *requests* an overview drawing which shows the two components from which the distance could be determined. The engineering department on the other hand might *offer* only drawing sets for each component, but not an overview of the whole beam-line. In this case, the information request cannot be fulfilled, although the information need could have been satisfied by a simple measurement on location.

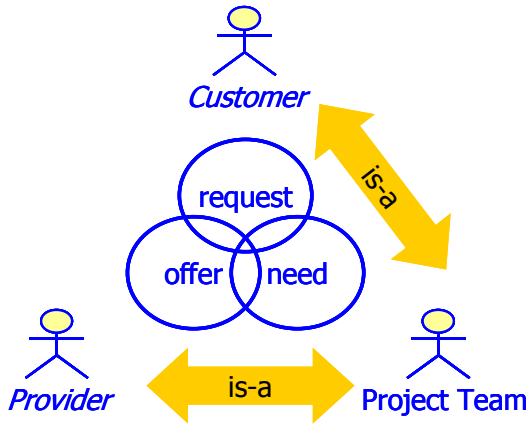

Figure 1. Adjusting information need, request and offer.

Information systems are the major tools for information management. The term usually refers to an IT application in conjunction with its integration into an information management process [5], and is differentiated to IT applications, which usually ignore the integration into its user's organization. To exemplify this definition, office suites are understood as IT applications, as it is the users' responsibility to adapt their way of working to the system, while e.g. ERP systems are information systems, since they can only be operated reasonably if they are used according to pre-defined use cases. In summary:

**Information management**
- ensures the information supply within project teams,
- adjusts the information which is requested, offered, and needed
- organizes information flow and the corresponding business processes, and
- introduces and operates information systems.

**Information systems**
- are defined as the entire infrastructure, organization, personnel,
- and components that collect, process, store, transmit, display, disseminate, and act on information [5].

**2.2. Collaboration Models.**

Understanding the philosophy of an organization's working model is an important precondition for information management, as the introduction of information systems interacts strongly with the organization.

Figure 2 shows two different collaboration models, which are frequently encountered in large organizations.

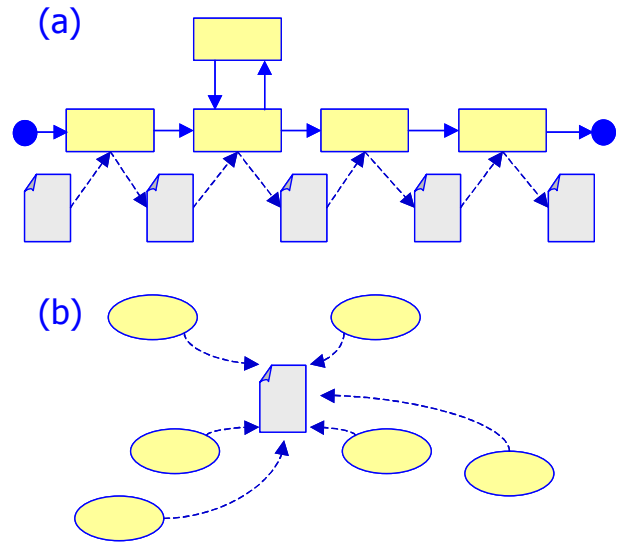

Figure 2. Different collaboration models and their implementation: (a) transaction processing, (b) marketplace.

Process-oriented organizations (Figure 2a) rely on well-defined and established processes, which are initiated by request or contract. Work packages and information flow are defined and known to the people involved. Processes are usually embedded in a strictly hierarchical organization. The work is predictable and calculable, individual steps can be optimized, the scenario is stable to change. Staff fluctuations affect the process only locally, as work packages are usually assigned to small teams or departments, and can thus be controlled. Similarly, quality control can be performed along the process, interventions are possible at early stages. Process-orientation is frequently seen in banking, insurance or manufacturing. Organizations with a large amount of standardization or routine work.

The working model of process-oriented companies is technically a transaction model, as it is similar to the architecture of e.g. database systems and hence maps well to information systems. Several straight-forward methods available for introducing information systems into process-oriented companies, which can easily scale to company-wide solutions for e.g. enterprise resource planning (ERP), production planning systems (PPS) or product lifecycle management (PLM).

An alternative to process-orientation is the collaboration model, which is typical for scientific communities. It is founded on the common interest of the participants in the intended result. Contributors self-organize around the subject of a project (Figure 2b). Every unit contributes according to its understanding and its capabilities. Work flow and information flow are coordinated through informal ad-hoc processes, which are initiated by common understanding. In this sense, collaboration can be viewed as a network of loosely-coupled expert groups.

Collaborations are less regulated than hierarchical process-oriented organizations and often achieve higher motivation of its contributors.

Technically, the working model of collaborations can be called a marketplace model. Marketplace models also occur in auctions and trading and B2B-applications. They are inherently distributed architectures which are sometimes implemented on top of groupware technology, but using groupware or similar approaches often restricts workflow management and process coordination to local tasks. This implies that overall project coordination and quality assurance are much harder than in transaction models, if not at all impossible.

### 2.3. Information Management for HEP.

The preparations for a global accelerator network (GAN, e.g. [6][7]) indicate that information management will be one of the key challenges of future accelerator projects. But HEP has a different culture of information exchange than industry, thus solutions from industry cannot simply be transferred to HEP.

Currently, collaborations in High Energy Physics are dominantly organized according to the marketplace model. On the other hand, building large accelerators and experiments are industrial scale projects of plant engineering and construction which seem to better fit into a process-oriented transaction-like model. The most critical areas are the engineering divisions, who are usually positioned on the boundary between innovative design work on the one hand, and routine quality production on the other hand. They thus need to implement both models, as for innovation marketplace models are beneficial, while design and production is better performed in transaction-like processing.

Thus information systems are tightly interwoven with the collaboration model. The request for novel information management tools implies therefore the necessity of organizational change. The attention will primarily focus on re-organizing the engineering processes. For the concerned groups, this implies a paradigm shift from the currently dominant collaboration model to a more strictly defined process-oriented way of work. This paradigm shift is often felt like moving the basic coordination principle from common understanding to work orders. Thus there will be inevitably some opposition, which has to be overcome. This calls for a dedicated acceptance management demonstrating the use and necessity of information management. Furthermore a strong management support is needed.

## 3. LIFE CYCLE SUPPORT

### 3.1. Life Cycle and Information Objects.

The life cycle of an accelerator can be coarsely structured into the four phases
- Envisioning and specification,
- design,
- production, i.e. manufacturing and installation, and
- operation and maintenance.

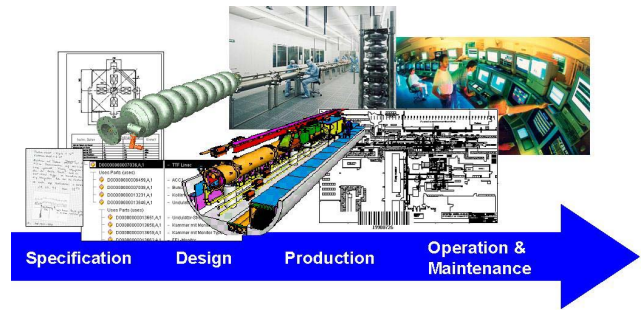

Figure 3. Life cycle and information objects.

Figure 3 illustrates the life cycle and shows that at each stage, characteristic information is emerging which needs to be collected for use in later phases. Envisioning predominantly results in specifications, sketches, calculations and presentations, while models, drawings and (assembly breakdown) structures are typical design entities. During production, information about individual parts is created, like e.g. work logs or observed manufacturing tolerances, followed by e.g. alignment information, maintenance records and other monitoring information which is collected during operation and maintenance.

It is an important observation that the entire information is needed as the life cycle proceeds. For example, manufactured parts have to be quality-controlled prior to installation. For this purpose, they have to be compared and related with specifications and design drawings. Design and specification information is also frequently requested when components have to be maintained.

### 3.2. Information Management Disciplines.

Each stage of the life cycle is characterized by different information entities and different user groups, which are addressed by different disciplines of information management. Major disciplines include
- Requirements Management (RM) addresses the elicitation and documentation of requirements on "products" (e.g. components),
- Product Data Management (PDM) coordinates the development and production of (accelerator, experiment etc.) components, and also covers Document Management,
- Asset Management (AM) handles purchasing, commissioning, operating and maintaining equipment and technical infrastructure, and
- Facility Management (FM) supports the planning, construction and maintenance of buildings and their technical installations, as well as realties.

Information management includes many more disciplines, among them project management and enterprise resource planning, but as this paper concentrates on technical engineering support, these topics are not discussed here any further.

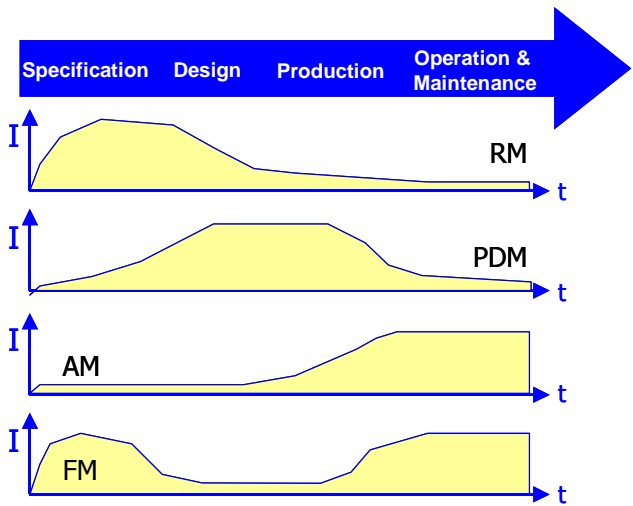

Figure 4. Different disciplines of information management with different intensities at different life cycle stages.

Figure 4 illustrates that the different disciplines contribute with varying intensities to the different life cycle stages. RM is essential at project initiation when ideas and constraints are collected and the intended components are defined. PDM is contributing during design and production, when engineering data is being generated, reviewed, approved etc., and when different engineering processes need to be coordinated. AM is gaining importance once production started to collect and store the technical parameters of each individual component, and FM is usually needed at project initiation to assess available grounds and buildings, and then as installation procedures commence. The relevance of the disciplines to HEP projects is discussed in more detail in section 4.

### 3.3. Mission.

To enable efficient information management throughout the life cycle, tools have to be provided which support the different management disciplines, and the engineering processes and the tools need to be adapted. This summarizes the mission of DESY's information management activities:
- provide optimum systems for each life cycle stage, and
- help to establish engineering processes which are compatible with the information management goals.

A major challenge is to differentiate the management discipline and the information system. Usually names are used synonymously for both (Figure 5), but the former encompasses process definitions and quality assurance, among other, while the latter refers to the technical solution: For example, Product Data Management (PDM) has to establish procedures for reviewing and releasing drawings, while a Product Data Management System (PDMS) offers check-out and check-in functionality for drawing files as a technical basis for implementing and supporting the procedures. Thus strictly speaking, PDM and PDMS are separate topics which require different responsibilities – while at the same time one cannot be treated without the other.

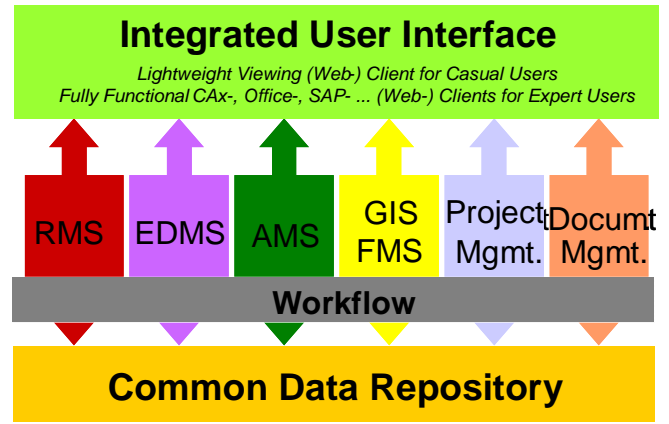

Figure 5. Management disciplines and information systems.

### 3.4. Integrated Information Management.

Information systems for supporting RM, PDM, AM and FM are required in several industries, however the combination of these systems as discussed in the previous subsections is rather uncommon. Aerospace industries, for example, need RMS, PDMS and AMS to design aircrafts and keep track of the parts, but no FMS. Railways need FMS and AMS to manage their tracks, stations and fleet, but no PDMS unless they build the trains themselves. It is thus possible to acquire commercial-of-the-shelf systems for supporting the different disciplines, but presently a solution which could support the entire life cycle in the above defined sense was unavailable.

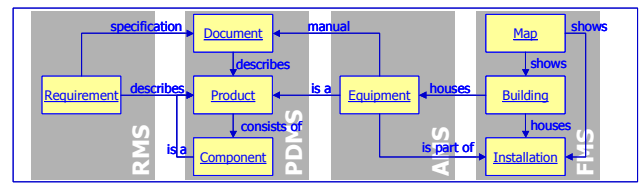

Figure 6. Information objects and their allocation to information systems.

Figure 6 shows a simplified schema which contains some of the information objects which were introduced in section 3.2, and illustrates how they are allocated to different information systems. The schema is separated because of the limited scopes of the information systems, and the repositories of the information systems have to be integrated to enable establishing all the relations which are shown in the diagram. Similar effects are occurring at every system layer, hence integration is also required at user interface and at kernel level. On the other hand, commercially available systems usually address a broader scope than only their core purpose. This can lead to the same functionality being available in more than one system.

## 4. INFORMATION MANAGEMENT AT DESY

Introducing information systems requires substantial lead time, as the systems have to be implemented and fully customized before an accelerator or an experiment can use the system for a given life cycle stage. Unfortunately lead times can easily amount to several years. Integrating information systems is yet another task, which is ideally planned already when introducing the first system, but which continues long after the systems were rolled out.

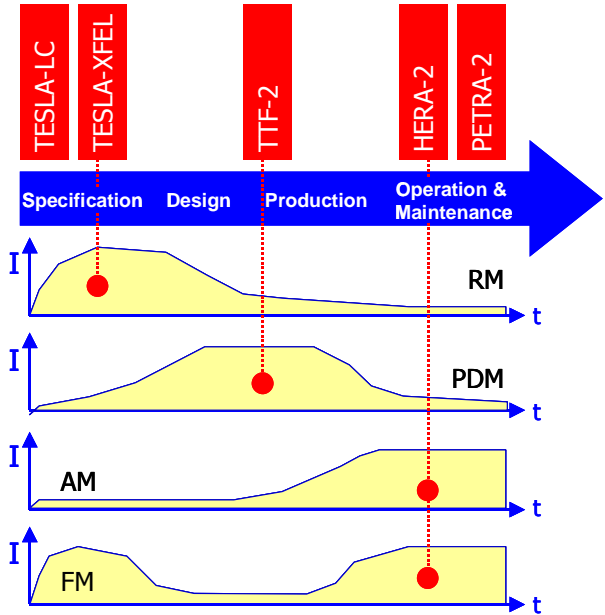

Figure 7. Accelerators at DESY and their life cycle stage.

DESY decided to introduce basic versions of the intended information systems in parallel and then extend them in collaboration with user groups. DESY is operating several accelerators which are at different stages in their life cycle, and the intention is to introduce every information system with focus on an accelerator which is in a corresponding life cycle stage.

Figure 7 positions some of DESY's accelerators along the life cycle. The chosen constellations for introducing information systems are

- a Requirements Management System (RMS) for TESLA planning,
- a Product Data Management System (PDMS) for supporting the engineering and the cavity preparation procedures of TTF-2,
- an Asset Management System (AMS) for managing components at HERA-2 and PETRA-2, and
- a Facility Management System (FMS) and Geographic Information System (GIS) for TESLA planning and "the campus".

The following sections describe these projects.

### 4.1. Requirements Management.

In preparation for the planned linear collider TESLA-LC and the free electron laser TESLA-XFEL, DESY began planning and designing the required buildings and facilities for the accelerators and experiments. The design team was formed of several experts with different backgrounds, including accelerator physics and research, engineering, safety, survey, cryogenics, facilities and many more.

The experts formed several loosely-coupled special interest groups which started specification and design work. The expert groups played the role of "stakeholders", who specify their ideas in terms of requirements and constraints, and civil engineers were in parallel developing designs of the buildings and facilities according to these specifications. To account for the technical dependencies of the different buildings, facilities and components, higher-level collaboration tools and methods were demanded.

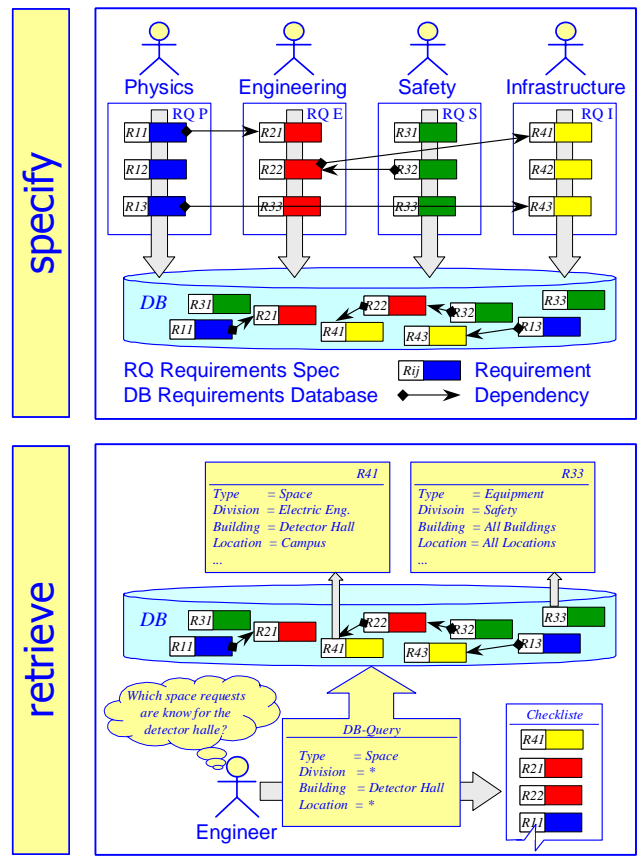

Figure 8. RMS solution for TESLA planning.

A solution was built around a commercial Requirements Management System (RMS). The RMS enabled the expert groups to create their specifications using their accustomed office tools. The RMS was then replicating individual paragraphs from the specification in a requirements database. One person per expert group was responsible for classifying the requirements in the database according to pre-defined categories. This way, the civil engineers were able to retrieve from the database all the requirements which are relevant for a given facility or building. The RMS includes a Web client which offers access to both the specifications and the requirements database. Figure 8 summarizes the RMS solution.

The RMS yielded two major benefits:
- The RMS enabled the "negotiation" of requirements, i.e. it helped to discover topics which have been addressed independently by more than one expert group and which needed to be synchronized.
- The RMS provided the essential checklists for approving the designs from the civil engineers.

The RMS has been successfully introduced to the TESLA planning group and is now containing the official specifications of several experts groups. The RMS became available only after the planning activities had already started, so it was adopted only reluctantly by the expert groups. Additionally, requirements management was a rather unusual activity for most of the groups, so substantial initial efforts went into adapting and training an RM method to the planning group. An important and fruitful measure was to introduce a (part-time) requirements manager into the project coordination team [8][9].

### 4.2. Product Data Management.

The components of the TTF-2 accelerator are designed, manufactured and commissioned at distributed locations. Within this framework, the cavity preparation is an important procedure which ensures and improves the quality of the accelerating cavities prior to their assembly in the accelerator modules. While the modules are being assembled and tested, ongoing research efforts at the same time continue pushing forward the cavity performance as long as possible.

A commercial PDMS is in production, which is used to coordinate the cavity preparation process, and to manage all the technical documentation which occurs in this process. Forty-two work packages have been defined which are controlled by a dynamic work flow which coordinates, traces and archives all steps. The PDMS treats cavities as parts, and for every produced individual new cavity, a corresponding new seralized part has to be created in the PDMS. The PDMS then automatically generates a new set of work packages for that individual cavity. The preparation process is iinciated when the process manger sends the first work package to the receiving team. Once a work package is finished, the owning team has to approve the work in the PDMS, which then notifies the entire teams and issues the next work package.

The cavity preparation process has been automated and simplified with the introduction of the PDMS, as notification and information exchange have been implemented electronically. Furthermore, the transparency of the process was increased, and the history of each cavity is now available from a central information point. But most important, due to the PDMS introduction, the preparation process needed to be defined, triggering the ongoing optimization of the process.

Figure 9 summarizes the influence of the PDMS introduction on the cavity preparation process.

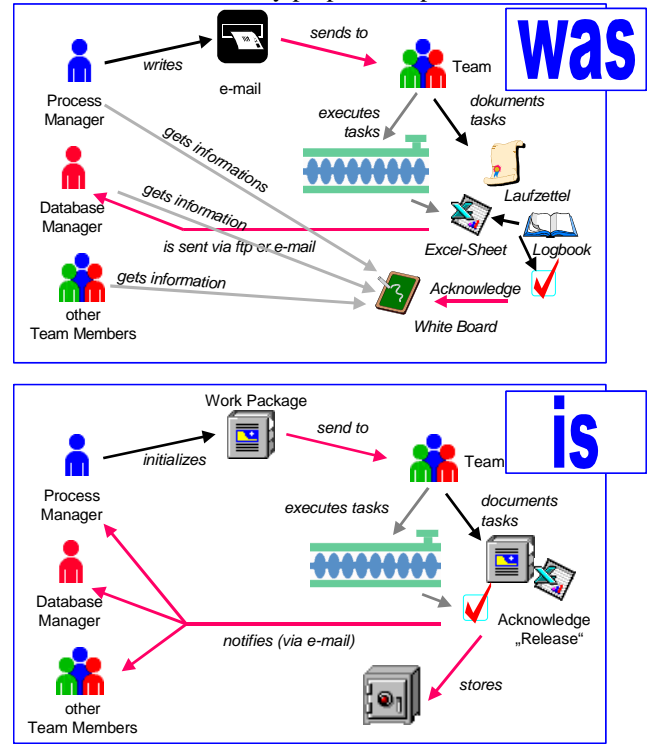

Figure 9. Cavity preparation process optimization due to PDMS introduction.

The PDMS is also used in the general engineering procedures of the TTF-2 collaboration. Users create part structures for components under development, using e.g. part-to-part relations for building a part breakdown structure (PBS). Then, documents like e.g. CAD models and drawings, work packages or general documentation are attached to parts, and the workflow engine coordinates the approval and notification procedures. All PDMS accesses are by means of a Web client. Figure 10 illustrates the two different structures which are kept in a PDMS.

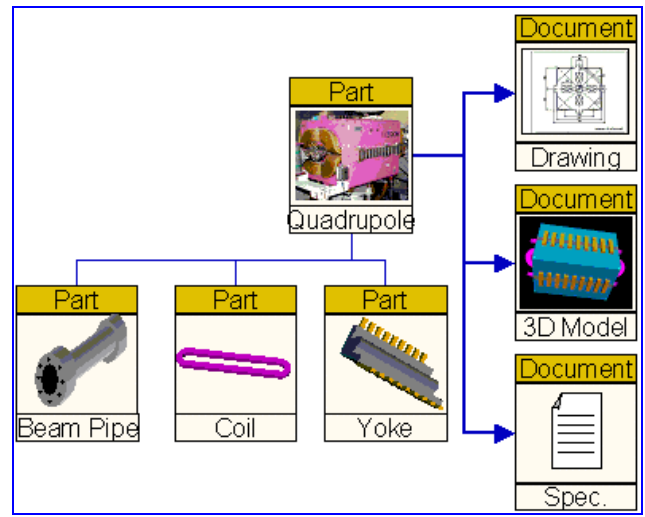

Figure 10. Part and document structures in a PDMS.

The demand for a PDMS and especially for document management is also known at other HEP laboratories, but implementing a PDM solution is always a tedious task. A PDMS is an highly complex information system, which first has to be carefully customized to be compatible with the laboratory's documentation standards and quality assurance procedures, and which after that requires a high degree of discipline of its users, which have to work according to these standards and procedures.

Prior to the PDMS introduction, a Web-based Document Management System (DMS) which has been developed in the HEP community was introduced at DESY [10]. The system had at that time only a basic and limited functionality. Using the system was an important step towards understanding the potential of a PDMS. The DMS enabled the users to explore the potential of document management, and when reaching the limitations of the available solution, to naturally develop the requirements for a PDMS.

### 4.3. Asset Management.

The HERA and PETRA accelerators are in operation for far more than a decade, and the major tasks at these machines are maintenance and upgrade work. The scope ranges from exchanging components or parts of components, like e.g. magnets, coils of magnets of vacuum pumps, to re-engineering whole accelerator sections

An AMS is being introduced to offer a single point of access to the accelerator "inventory", and to improve spare parts management. The AMS contains a central repository for equipment, which stores technical and operational parameters and demographic information for every individual component and keeps track of their history. A workflow engine supports moving, adding and changing components, and it can be used to automate e.g. the purchasing of spare parts, or to issue and track work orders for maintenance tasks. The AMS is accessed through a Web client.

An first version of the AMS is containing the inventory data of the accelerators and is ready for roll-out. The AMS will be initially used by the group who is responsible for coordinating the maintenance work at the accelerators.

Figure 11. Technical information in an AMS.

The selected AMS is a general-purpose system which is open to accommodate technical infrastructure of many different categories. DESY has decided to introduce the AMS first for the IT infrastructure, and to extend it to other domains after initial experience has been gained. From a process-oriented point of view, the system requirements for supporting operation and maintenance of IT hardware and of accelerator components are similar to a certain degree, hence the AMS is a promising candidate for generating synergy. The IT-AMS is in production [11].

### 4.4. Facility Management.

The upcoming and planned future accelerator projects initiate extensive civil engineering activities. A facility management solution is being prepared to integrate the planning and documentation activities of civil engineering, accelerator design and installation of technical infrastructure.

The intended solution is based on three commercial products: An architectural CAD system is the basis for creating plans of buildings and halls. The plans are structured into layers, and every user group (e.g. civil engineers, safety personnel or electricity providers) maintains a set of layers in the common plans.

A FMS contains a central repository for building information and enables e.g. user, key or maintenance management. It features an online connection with the CAD system, enabling navigation between CAD plans and the FM-database.

The solution is completed by a GIS, which maintains maps of the campus and the accelerator tracks and is the basis for survey and alignment.

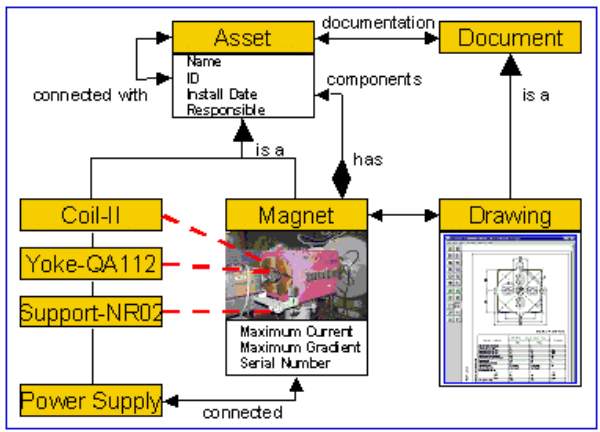

**TUNP002**

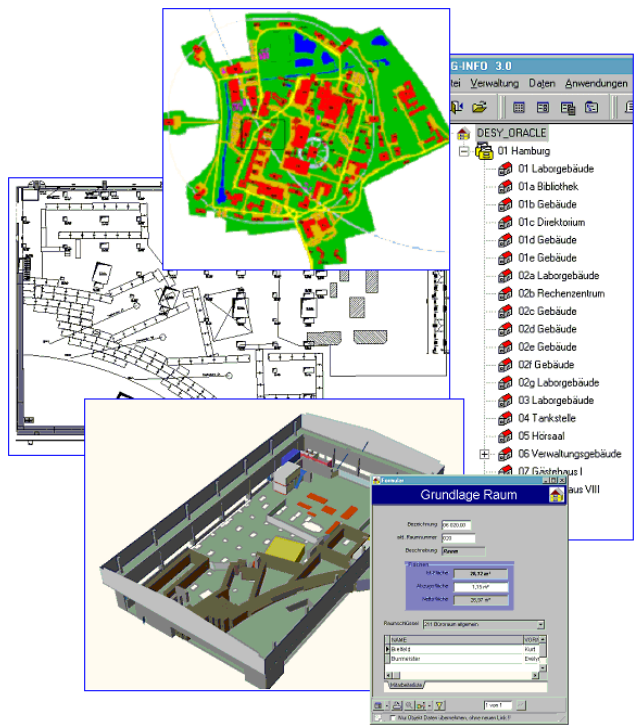

Figure 12. Components of a facility management solution

Figure 12 shows some components of the solution. Currently, a legacy GIS is in production, and a pilot for demonstrating the intended new GIS/FMS-solution is available.

### 4.5. Integration Efforts.

Once the information systems are in production, efforts have to be started to provide the intended integrated solution. Conceptually, the systems should (at least to the casual user) appear as one information management platform.

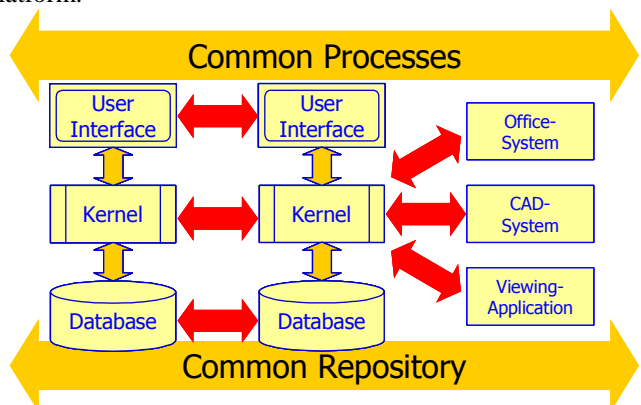

Figure 13. Integration activities.

Figure 13 illustrates the integration efforts which are required for providing such a platform. The information systems need to be integrated at every layer, i.e. the repository schemas need to be connected, transactions have to be propagated among the systems, an the user interfaces have to enable navigation across system boundaries. Additionally, specialized applications like e.g.

CAD-systems, office-suites or third-party databases have to be integrated tightly with a given information systems, so that information created within these applications is automatically propagated into the information management platform. These type of integration is often realized commercially by the vendors of the information systems.

Several concepts are existing for system integration [12], but currently no generic method for system integration is known in computer science.

## 5. RESULTS AND EXPERIENCE

### 5.1. Results.

Most of the intended information systems are in production:
- The RMS is in production since Q3/2002 and is used for the TESLA planning activities.
- The PDMS is in production since Q4/2001 and is used for the cavity preparation and the engineering of TTF-2, and it contains the documentation of the TESLA planning activities.
- The AMS is in production for IT equipment since Q3/2002, and is ready for an initial roll-out in the accelerator domain.
- The GIS/FMS solution is available for demonstration, and an initial roll-out is scheduled for the end of 2003.

In addition to system introduction, initial integration efforts have been started.

Choosing different target groups enabled a parallel introduction of the different information systems, which is considered necessary because of the large lead time which a system introduction needs. As a drawback, system integration becomes much harder, as there are almost no common objects in the different repositories.

### 5.2. Observed Benefits.

In their first year of operation, the systems have been able to demonstrate some of the benefits of information management:

**The systems act as a central communication and documentation platform**, offering a single point of access to information from a heterogeneous environment. All users have structured access to up-to-date information, which can originate from a variety of different sources (e.g. Office, CAx). The systems enable associating and querying such information, and they help to discover gaps or conflicts in the existing information pool. The latter has been observed especially with the RMS.

**The systems enable and improve concurrent engineering**, which has been observed in the TTF-2 engineering processes and in the TESLA planning activities. Information becomes available to independently operating distributed working groups, and information from different engineering domains is integrated.

**The systems help to improve the (business) processes**

**TUNP002**

The information systems enforces the user groups to (re)define working procedures. If a not optimized process is supported by an information system, it remains a not optimized process. The fact that the process has to be defined makes deficits visible and leads to re-engineering and optimizing the process. This has been especially observed for the cavity preparation process.

**The systems ensure long term archiving of emerging knowledge**. They store documents and information both in their native formats and in a neutral presentation format (e.g. pdf). The later ensures that they can be accessed even after long periods of time. The documents can be modified as long as the native application is supported. This feature is essential for interrupting and resuming activities, which occurs frequently when e.g. projects are waiting for approval and funding.

### 5.3. Experience.

From a technical point of view, introducing information management "simply" requires, that the entire life cycles become coordinated with the business processes. Methods are known for implementing such solutions in process-oriented organizations.

In collaborations, however, methods and solutions are less formalized and thus leave a higher degree of uncertainty whether they will finally succeed. Experience shows that the key success factors for introducing information management do not include technical issues. The following paragraphs summarize parts of the experience gained.

**The major success factors** of information management in a HEP environment are use, usability and extensive user support. Any information system (or process optimization) has to deliver immediate and transparent benefit to the user, otherwise it will not be adopted voluntarily. The latter implies that the information pool will remain incomplete and that therefore the intended project goals can hardly be achieved, unless system usage is enforced. Usability is the means by which reluctant users can be convinced to join the community, and user support is necessary to provide training in method and tools.

**Be patient and be quick.** Information management projects need long advance planning and lead times, and once a system has been successfully rolled-out, it takes equally long until it is adopted. It is important to communicate these timing constraints to users and management, but at the same time develop a strategy for incremental delivery to be prepared for continuously demonstrating the progress of the project.

**The tool is always guilty.** Introducing document-based information systems like PDMS or RMS is especially tedious, because as long as the users don't create documents within the systems, the systems simply stay empty and useless, even if they offer their full technological potential. Users tend to not differentiate between tasks, methods and tools, but rather assign any property to the tool. Thus a lack of documentation, which indicates an organizational problem, is quickly perceived as being due to a bad system. Figure 14 illustrates the effect.

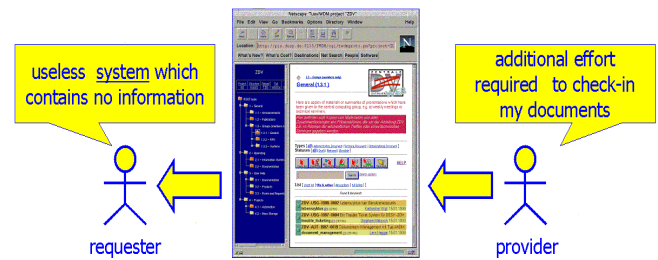

Figure 14. Barriers for document-based information systems.

An inefficient, but sometimes the only countermeasure is to foresee that the information management team joins the user groups and performs their documentation tasks until the system usage is established.

In case of database-oriented systems the situation is a little more relaxed as usually some legacy data is available which can be imported by the project team to initialize system usage.

**Turn post-mortem acceptance into promotion.** The value of an information system becomes visible when it is threatened to be retired. At this point, the legacy data is scanned, and the users communicate the drawbacks and problems they will experience if the system became unavailable: they will concentrate on the system strengths instead of its weaknesses. It is important for the success of a project, that the strengths of the solution are stressed from the very beginning. Asking users early what they would loose if the system were retired is an efficient way to discover what strengths are felt by the user community, and to sensitize the users to the benefits they are receiving.

**Beware the integration.** Integration is a difficult project of its own, and to add difficulties, integration tasks are completely invisible to users and management and thus hard to be equipped with resources and budget. The conceptual integration work has to be started very early, while the technical work may be started only after initial versions of the information systems are available. Some conceptual issues which need to be taken into account early include system architecture, access and security policies, and data schema and data exchange.

**The "Demonstrate-the-Benefits" Dilemma.** For existing accelerators, information is sometimes only available as paper-based documentation, and the status and location of the documents are maintained only locally by the authoring group. This may become problematic when e.g. after years of operation, accelerators or experiments have to be modified (e.g. for upgrades), and the authoring experts have in the mean time become unavailable. These difficulties can only be prevented, if emphasis is put on creating and maintaining information already during the very first stages of the life cycle, leading to the dilemma that

- ensuring efficient long-term maintainability requires extensive and thorough documentation during early life cycle stages,
- documentation is felt as an additional and non-productive workload,
- the benefits of documentation become visible only in later stages of the life cycles, and only after extensive documentation has been created, and
- the benefiting groups are usually different from the groups which create the documentation.

To resolve this dilemma is one of the major challenges of information management.

## References.